\documentclass[aps,prl,twocolumn,showcaps,superscriptaddress,showpacs,amsmath,amssymb]{revtex4}
\usepackage{graphicx}

\newcommand{\DD}{\mathrm{D}}
\newcommand{\dd}{\mathrm{d}}

\newcommand{\mean}[1]{\langle #1 \rangle}

\newcommand{\mat}[1]{\mathbf #1}

\renewcommand{\vec}[1]{\mathbf #1}
\DeclareMathOperator{\tr}{tr}

\newcommand{\al}{\alpha}

\newcommand{\lam}{\lambda}
\newcommand{\gam}{\gamma}

\newcommand{\id}{\mathbf 1}     

\newcommand{\X}{\Gamma}         
\newcommand{\x}{\vec r}         
\newcommand{\vl}{\vec v}        

\newcommand{\F}{\boldsymbol{\mathcal{F}}}
\newcommand{\mo}{\boldsymbol{\mu}}
\newcommand{\kap}{\boldsymbol{\kappa}}
\newcommand{\eq}{_\mathrm{eq}}  
\newcommand{\T}{^\mathrm{T}}    
\newcommand{\sym}{^\mathrm{S}}  
\newcommand{\asy}{^\mathrm{A}}  
\newcommand{\med}{_\mathrm{m}}  
\newcommand{\tot}{_\mathrm{tot}} 

\newcommand{\shr}{\tilde\gamma}
\newcommand{\altn}{^\ast}

\begin{document}


\title{Role of External Flow and Frame Invariance in Stochastic
  Thermodynamics}

\author{Thomas Speck}
\author{Jakob Mehl}
\author{Udo Seifert}

\affiliation{{II.} Institut f\"ur Theoretische Physik,
Universit\"at Stuttgart, Pfaffenwaldring 57, 70550 Stuttgart,
Germany}

\begin{abstract}
  For configurational changes of soft matter systems affected or caused by
  external hydrodynamic flow, we identify applied work, exchanged heat, and
  entropy change on the level of a single trajectory.  These expressions
  guarantee invariance of stochastic thermodynamics under a change of frame of
  reference. As criterion for equilibrium \textit{vs.} nonequilibrium, zero
  \textit{vs.} nonzero applied work replaces detailed balance \textit{vs.}
  nonvanishing currents, since both latter criteria are shown to depend on the
  frame of reference. Our results are illustrated quantitatively by
  calculating the large deviation function for the entropy production of a
  dumbbell in shear flow.
\end{abstract}


\maketitle


Thermodynamic notions like applied work, dissipated heat, and entropy have
been used successfully to analyze processes in which single colloidal
particles or biomolecules are manipulated externally~\cite{bust05}. Various
exact relations have been shown to constrain the distribution function arising
from the ever present thermal fluctuations on this
scale~\cite{jarz97,croo99,hata01,seif05a}. So far, the systematic conceptual
work has been focused on cases where the driving crucial to generate a
nonequilibrium situation arises from a time-dependent potential expressing the
effect of a moving laser trap, micropipet, or AFM tip. In these cases, the
identification of external work, internal energy and hence dissipated heat is
straightforward. Many experiments on a variety of systems have proven the
consistency and potency of such an extension of thermodynamic notions to the
micro or nano world~\cite{wang02,liph02,trep04,doua06,blic06,coll05}.

As another source of nonequilibrium, external hydrodynamic flow arguably is
the most common and best studied case in soft matter systems. On the level of
single objects like a polymer, vesicle, or red-blood cell, it can cause
dramatic shape transitions (for recent examples see, e.g.,
Refs.~\cite{wink06,misb06,nogu07,abka07,kant06}). The theoretical analysis of
these phenomena is typically based on equations of motion like the Langevin
equation for single objects or distribution functions and their projection to
few-body correlation functions for bulk systems like colloidal
suspensions~\cite{doiedwards,dhont,brad07}. Concepts like work or entropy
production, however, have played no systematic role in describing such
phenomena yet. Having in mind the conceptual advantage achieved for
time-dependent potentials by using such notions, the question arises quite
naturally whether a similar analysis for nonequilibrium phenomena in soft
matter systems caused by external flow is possible. Indeed, Turitsyn
\textit{et al.}~\cite{turi07} calculated the entropy production for linear
equations of motion and illustrated it for a dumbbell in shear flow. However,
their identification of entropy production was not related to the dissipated
heat leading to a fundamentally different dependence on the external flow.

Conceptually, there is an even deeper issue hidden behind the proper treatment
of external flow. Its presence or absence depends on the frame of reference as
the presence or absence of currents does. Since the latter are usually taken
as an indicator for nonequilibrium, our analysis will question the traditional
role of both detailed balance and nonvanishing currents as criteria for
equilibrium and nonequilibrium, respectively. They will be replaced by a
frame-independent identification of applied work.


For an almost trivial but revealing paradigmatic case, consider a colloidal
particle dragged through a viscous fluid along the trajectory $x(t)$ by a
harmonic laser trap of strength $k$ moving with velocity
$u_0$~\cite{wang02,trep04}. In the laboratory frame, the particle is moving in
a time-dependent potential $U(x,t)=(k/2)(x-u_0t)^2$. By applying the standard
definition of work~\cite{jarz97} as the external change of the potential
energy, the applied power
\begin{equation}
  \label{eq:w}
  \dot w = \partial_t U
\end{equation}
becomes $\dot w=-u_0k(x-u_0t)$. Changing to the comoving frame with $y\equiv
x-u_0t$, the potential $U(y)=(k/2)y^2$ becomes time-independent and therefore
one would naively find $\dot w=0$. Of course, the work should be independent
of the chosen frame of reference. Moreover, since in the comoving frame
detailed balance holds and no current in the $y$ variable occurs, one would
usually consider the system to be in equilibrium in this representation. The
physical origin of the apparent contradictions is the fact that in the
comoving frame the particle experiences a steady flow of the fluid with
velocity $-u_0$. Due to friction, this flow pushes the particle against the
force of the potential thus spending work. The standard definition of work
fails in the presence of flow since the flow advects the particle, which is
not accounted for in Eq.~\eqref{eq:w}. Modifying this expression will affect
also the expression for the heat dissipated into the fluid as well as the
expression for the entropy production.


We will now derive the refined expressions for work and heat in the presence
of a, possibly time-dependent, external flow $\vec u(\x,t)$. For complete
generality and future applications, we consider a soft matter system like
polymers, membranes, or a colloidal suspension composed of $N$ particles with
positions $\X\equiv\{\x_1,\dots,\x_N\}$. The index denotes the particle number
and in the following we sum over same indices. The system has a total energy
$U(\X,t)$ which is the sum of an internal energy due to particle interactions
and a possible contribution due to external potentials. In addition to direct
interactions, we allow for hydrodynamic interactions which are of great
importance for soft matter systems.

In 1997, Sekimoto formulated the first law of thermodynamics $\delta q=\delta
w-\dd U$ along a single stochastic trajectory~\cite{seki97}. We follow this
route and start by identifying three possible sources of work $w$, i.e., of
changes of energy caused externally. First, the potential energy $U(\X,t)$ can
be time-dependent if it is manipulated externally. Second, it is convenient to
allow for direct nonconservative forces $\vec f_k$ that spend work through
displacing the particles. Third, the external flow drives the system and
therefore changes its energy, which must be taken into account as work as
well. The advection of particles is described through the convective
derivative $\DD_t\equiv\partial_t+\vec u(\x_k)\cdot\nabla_k$. Through both
replacing the partial derivative in Eq.~\eqref{eq:w} with the convective
derivative and measuring particle displacements with respect to the flow, we
obtain the new definition
\begin{equation}
  \label{eq:work}
  \dot w \equiv \DD_tU + \vec f_k\cdot[\dot\x_k-\vec u(\x_k)]
\end{equation}
for the work increment $\delta w=\dot w\dd t$ which replaces Eq.~\eqref{eq:w}.
The first law then leads to the heat production rate
\begin{equation}
  \label{eq:heat}
  \dot q \equiv \dot w - \dd U/\dd t
  = [\dot\x_k-\vec u(\x_k)]\cdot[-\nabla_kU + \vec f_k].
\end{equation}
The sign of heat is convention, here we take it to be positive if energy is
dissipated into the surrounding fluid. The expression~\eqref{eq:heat} shows
that no heat is dissipated when particles move along a fluid trajectory,
$\dot\x_k=\vec u(\x_k)$. In this case, the applied work corresponds to the
change in total energy.

A change of frame necessarily changes the flow therefore leaving the
expressions for work~\eqref{eq:work} and heat~\eqref{eq:heat} invariant.
Explicitly, we can allow time-dependent orthogonal transformations and an
arbitrary time-dependent shift of the origin. Coming back to the introductory
example, we see that for a stationary trap in the comoving frame with flow
$u=-u_0$, the applied power following Eq.~\eqref{eq:work} now reads $\dot
w=-u_0ky$ in agreement with the rate obtained for a moving trap and resting
fluid. Finally, note that the definitions~\eqref{eq:work} and~\eqref{eq:heat}
are independent of whether or not hydrodynamic interactions are induced. These
will affect the dynamics but they do not enter the definitions of work and
heat explicitly.


We now turn to entropy production. A trajectory dependent stochastic entropy
is defined as $s(t)\equiv-\ln\psi(\X(t),t)$ through the distribution function
$\psi(\X,t)$ of the $N$ particle positions~\cite{seif05a}. The temperature of
the surrounding fluid is $T$ and throughout the paper, we set Boltzmann's
constant to unity. Calculating the total time derivative $\dot s\equiv\dd
s/\dd t$, we obtain the equation of motion
\begin{multline}
  \label{eq:s:dot}
  \dot s = \partial_t s - \dot\x_k\cdot\nabla_k\ln\psi
  = \DD_t s + [\dot\x_k-\vec u(\x_k)]\cdot\F_k/T \\
  - [\dot\x_k-\vec u(\x_k)]\cdot[-\nabla_kU + \vec f_k]/T,
\end{multline}
where we have separated the heat production rate~\eqref{eq:heat} and
introduced the total effective force
\begin{equation}
  \F_k \equiv -\nabla_k(U+T\ln\psi) + \vec f_k.
\end{equation}
Apart from the gradient of the potential energy $U(\X,t)$ and nonconservative
forces $\vec f_k$ that cannot be written as gradient of a potential, a
``thermodynamic'' force contributes to $\F_k$ arising from the stochastic
interactions between system and the surrounding fluid~\cite{doiedwards}. In
the absence of external flows and nonconservative forces, detailed balance
must hold. The thermodynamic force then ensures that $\nabla_k(U +
T\ln\psi)=0$ leads to the correct Gibbs-Boltzmann equilibrium distribution
$\psi\eq\sim\exp(-U/T)$. In Eq.~\eqref{eq:s:dot}, the last term is the
heat~\eqref{eq:heat} divided by the temperature of the fluid $T$. We interpret
this term as the entropy produced in the fluid through Clausius' formula
$\Delta s\med=q/T$. The total entropy production rate $\dot s\tot\equiv\dot
s+\dot s\med$ then becomes
\begin{equation}
  \label{eq:stot}
  \dot s\tot = \DD_ts + [\dot\x_k-\vec u(\x_k)]\cdot\F_k/T.
\end{equation}
The mean total entropy production rate follows as
\begin{equation}
  \label{eq:stot:mean}
  T\mean{\dot s\tot} = \mean{[\dot\x_k-\vec u(\x_k)]\cdot\F_k}
\end{equation}
since the mean of the convective derivative is zero for incompressible fluids
($\nabla\cdot\vec u=0$) and vanishing boundary terms.

So far we did not resort to a specific dynamics. However, in order to both
prove positivity of Eq.~\eqref{eq:stot:mean} and give it a more familiar
appearance known from the theory of polymer dynamics~\cite{doiedwards}, we
turn to the Smoluchowski equation~\cite{doiedwards,dhont}
\begin{equation}
  \label{eq:sm}
  \partial_t\psi + \nabla_k\cdot(\vl_k\psi) = 0
\end{equation}
governing the evolution of the distribution function $\psi(\X,t)$. Any
deviation of a particle's local mean velocity
\begin{equation}
  \label{eq:vloc}
  \vl_k \equiv \vec u(\x_k) + \mo_{kl}\F_l
\end{equation}
from the velocity of the external flow $\vec u(\x)$, which drags the particles
due to friction, has to be caused by the total force $\F_k$. In this model,
hydrodynamic interactions now enter through a dependence of the symmetric
mobility matrices $\mo_{kl}(\X)$ on the particle positions $\X$. We demand the
inverse matrices defined through $\mo_{km}\mo^{-1}_{ml}=\id\delta_{kl}$ to
exist. Specific expressions for $\mo_{kl}$ are obtained from either the Oseen
or Rotne-Prager tensor~\cite{dhont}.

The local mean velocity $\vl_k(\X,t)$ is the average of the actual stochastic
velocity $\dot\x_k$ over the subset of trajectories passing through a given
configuration $\X$. This allows us to perform the mean in
Eq.~\eqref{eq:stot:mean} in two steps: First, we average over stochastic
trajectories crossing a specific point $\X$ in configuration space, which
amounts to replacing $\dot\x_k$ by $\vl_k$. Second, we average over the
distribution $\psi(\X,t)$ of the point $\X$. Finally, we use
Eq.~\eqref{eq:vloc} to replace the total force $\F_k$. The resulting quadratic
expression
\begin{equation}
  T\mean{\dot s\tot} = \mean{[\vl_k-\vec u(\x_k)]\cdot\mo^{-1}_{kl}
    [\vl_l-\vec u(\x_l)]} \geqslant 0
\end{equation}
is well known phenomenologically as the energy dissipation function in the
theory of polymer dynamics~\cite{doiedwards}. It is nonnegative with the equal
sign holding in equilibrium only where $\F_k=0$.


On the trajectory level, we have identified heat and entropy production using
physical arguments. There is a more formal way of identifying entropy by
starting from the weight $P[\X(t)|\X_0]\sim\exp\{-S[\X(t)|\X_0]/T\}$ for a
trajectory $\X(t)$ starting in $\X(0)=\X_0$ involving the action $S$. The
entropy produced in the surrounding medium is usually identified with the part
of the action that is asymmetric with respect to
time-reversal~\cite{maes03}. For explicit expressions for both the path
probability and the action in the presence of hydrodynamic interactions as
needed in our case, we refer to Ref.~\cite{maes07a}.

We can deduce two expressions for the asymmetric part of the action since in
the presence of external flow, we have two choices how to define the operation
``time reversal''.  First, we could formally treat the flow as a
nonconservative force.  It would then be invariant with respect to time
reversal and we would obtain
\begin{equation}
  \label{eq:asym:1}
  \dot s\med\altn
  = \dot\x_k\cdot[-\nabla_kU + \vec f_k + \mo^{-1}_{kl}\vec u(\x_l)]/T.
\end{equation}
This expression has been identified with the entropy production rate in
Ref.~\cite{turi07}. Second, we can extend the operation of time reversal to
the particles of the fluid, which is physically more appropriate. Reversing
the velocity of the fluid particles then effectively amounts to the change of
the sign of the external flow velocity $\vec u(\x)$. This leads to $\dot
s\med=\dot q/T$ involving the heat production rate from
Eq.~\eqref{eq:heat}. We thus recover the physically motivated definition of
the heat~\eqref{eq:heat} as the asymmetric part of the action under time
reversal only if the flow is reversed as well~\cite{turi07}.


For a discussion of the crucial difference between Eqs.~\eqref{eq:asym:1}
and~\eqref{eq:heat} in a specific system, consider a dumbbell in a flow $\vec
u(\x)=\kap\x$ with traceless matrix $\kap=\kap\sym+\kap\asy$, which can be
split up into a symmetric part $\kap\sym$ and a skew-symmetric part
$\kap\asy$. The dumbbell consists of two particles at positions $\x_1$ and
$\x_2$ with displacement $\x\equiv\x_1-\x_2$ and potential energy
$U(\x)=(k/2)\x^2$. The applied power~\eqref{eq:work} then becomes
\begin{equation}
  \label{eq:dumbbell}
  \dot w = \vec u(\x)\cdot\nabla U = k\x\cdot\kap\sym\x,
\end{equation}
i.e., only the symmetric part $\kap\sym$, which reflects an elongation flow
component, contributes to the work. Physically, the elongation flow drags the
two particles apart due to friction and therefore spends permanently work
against the elastic force keeping the particles together. Neglecting the
boundary term $\Delta U/t\sim 0$ in the long time limit, the time-extensive
part of the entropy production rate is then $\dot
s\med\sim(k/T)\x\cdot\kap\sym\x$. Note that here only the displacement $\x$
enters. This is not true if the expression~\eqref{eq:asym:1} is interpreted as
entropy production rate, leading to the time extensive part
\begin{equation}
  \label{eq:wrong}
  \dot s\med\altn \sim (T\mu_0)^{-1}[2\dot{\vec R}\cdot\kap\asy\vec R 
  + (1/2)\dot\x\cdot\kap\asy\x]
\end{equation}
with bare mobility $\mu_0$. Here, also the center of mass $\vec
R\equiv(\x_1+\x_2)/2$ contributes. This is obviously physically unsound as the
center of mass undergoes free diffusion and therefore does not produce entropy
in the medium on average. Moreover, even in the second term of
Eq.~\eqref{eq:wrong} concerning the relative motion, the source of dissipation
is the skew-symmetric part $\kap\asy$ in contrast to Eq.~\eqref{eq:dumbbell}.

We will now calculate the large deviation function of the medium entropy
production rate for the dumbbell in two-dimensional shear flow, i.e., all
entries of the matrix $\kap$ are now zero except $\kappa_{xy}=\dot\gam$ where
$\dot\gam$ is the strain rate~\cite{dhont}. The large deviation function
\begin{equation}
  \label{eq:ldf}
  h(\sigma) \equiv \lim_{t\rightarrow\infty}-\frac{1}{t}\ln p(\Delta s\med,t)
\end{equation}
quantifies the asymptotic fluctuations $\sigma=\Delta s\med/(\mean{\dot
  s\med}t)$ of the entropy production in the limit of large observation times
$t$ with mean production rate $\mean{\dot s\med}$. Instead of obtaining the
function $h(\sigma)$ from the time-dependent probability distribution
$p(\Delta s\med,t)$ directly, we will calculate its Legendre transform
$\al_0(\lam)$ as the lowest eigenvalue of the operator $\hat L_\lam=\hat
L_0-\lam\dot s\med$. The operator $\hat L_\lam$ governs the evolution of the
generating function~\cite{lebo99,visc06,impa07}. The production rate depends
only on the displacement $\x$, whose dynamics is determined through the
Smoluchowski operator~\cite{doiedwards} $\hat
L_0=\tau_0^{-1}\nabla\cdot[\x+(T/k)\nabla]-\kap\x\cdot\nabla$. For this
illustration, we neglect hydrodynamic interactions by using mobility matrices
$\mo_{kl}=\mu_0\id\delta_{kl}$ defining the time scale
$\tau_0^{-1}\equiv2\mu_0k$.

\begin{figure}[t]
  \includegraphics[width=\linewidth]{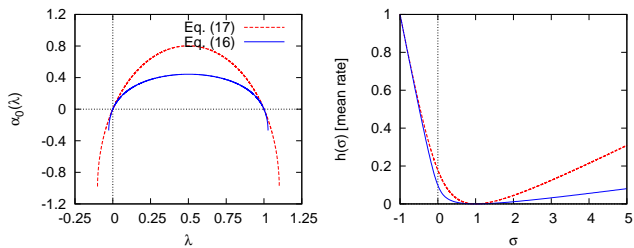}
  \caption{Left: Comparison of the eigenvalue $\al_0(\lam)$ from
    Eq.~\eqref{eq:sol} (solid) with the solution~\eqref{eq:tur} obtained in
    Ref.~\cite{turi07} (dashed) based on Eq.~\eqref{eq:asym:1}. (Parameters
    for both curves are $\shr=3$ and $\tau_0=1$.) The mean entropy production
    rate is the same for both expressions as indicated by the matching slope
    at $\lam=0$. Right: The corresponding large deviation functions
    $h(\sigma)$ obtained from the Legendre transform of $\al_0(\lam)$.}
  \label{fig:ldf}
\end{figure}

For the lowest eigenfunction of the eigenvalue equation
\begin{equation}
  \label{eq:eig}
  \hat L_\lam\psi_0(\x,\lam) = -\al_0(\lam)\psi_0(\x,\lam),
\end{equation}
we use the ansatz $\psi_0=\exp[-(k/2T)\x\cdot\mat C_\lam\x]$ with a symmetric
matrix $\mat C_\lam$. We are lead to this ansatz since we know that the
stationary distribution ($\lam=0$) of $\x$ is a Gaussian~\cite{john87}, where
$\mat C_0$ becomes the inverse covariance matrix. Inserting this ansatz
into~Eq.\eqref{eq:eig} results in $\al_0(\lam)=\tau_0^{-1}\tr(\mat
C_\lam-\id)$, where $\mat C_\lam$ is the solution of the quadratic matrix
equation
\begin{gather*}
  (\mat C_\lam+\mat D)\T(\mat C_\lam+\mat D) = \mat S_\lam, \\
  \mat S_\lam = \frac{1}{4}\left(
    \begin{array}{cc}
      1 & \shr(2\lam-1) \\
      \shr(2\lam-1) & 1+\shr^2
    \end{array}\right), \quad
  \mat D = \frac{1}{2}\left(
  \begin{array}{cc}
    -1 & \shr \\
    0 & -1
  \end{array}\right).
\end{gather*}
The only parameter left is the dimensionless strain rate
$\shr\equiv\dot\gam\tau_0$. We only need the trace of the matrix $\mat
C_\lam$, which, after some tedious calculations, can be expressed as
\begin{equation*}
  \tr\mat C_\lam = \sqrt{\tr\mat S_\lam+2\sqrt{\det\mat S_\lam}-\shr^2/4}
  - \tr\mat D
\end{equation*}
leading to the solution
\begin{equation}
  \label{eq:sol}
  \al_0(\lam) = \tau_0^{-1} \left[ \sqrt{\tfrac{1}{2}
      +\tfrac{1}{2} \sqrt{1+4\shr^2\lam(1-\lam)}} - 1 \right].
\end{equation}
This function is defined within the interval
$\lam_-\leqslant\lam\leqslant\lam_+$ with branch points
$\lam_\pm=\tfrac{1}{2}(1\pm\sqrt{1+\shr^{-2}})$. Such branch points imply
linear asymptotes for $h(\sigma)$ and therefore exponential tails for the
distribution $p(\Delta s\med)$~\footnote{An unbounded potential $U$ in
  connection with exponential tails of $p(\Delta s\med)$ might affect the
  large deviation function and lead to a modified fluctuation theorem, see
  Refs.~\cite{zon03,visc06}. A complete analysis of the present system in this
  regard will be published elsewhere.}. On the level of the function
$\al_0(\lam)$, the fluctuation theorem~\cite{evan93,gall95,lebo99,seif05a} is
expressed through the symmetry $\al_0(\lam)=\al_0(1-\lam)$, which is fulfilled
by Eq.~\eqref{eq:sol}. In Fig.~\ref{fig:ldf}, we compare the
function~\eqref{eq:sol} with the solution
\begin{equation}
  \label{eq:tur}
  \al\altn_0(\lam) 
  = \tau_0^{-1} \left[ \sqrt{1+\shr^2\lam(1-\lam)} - 1 \right]
\end{equation}
obtained in Ref.~\cite{turi07} based on the entropy production rate in
Eq.~\eqref{eq:wrong} with fixed center of mass ($\dot{\vec R}=0$), which also
fulfills a fluctuation theorem. Note that our expression for $h(\sigma)$ based
on Eq.~\eqref{eq:sol} predicts substantially larger fluctuations especially
for trajectories with larger than mean entropy production.

The dumbbell just discussed also provides a counter-example to the standard
definition of nonequilibrium. In pure rotational flow with skew-symmetric
matrix
\begin{equation*}
  \kap = \kap\asy = \left(
    \begin{array}{cc}
      0 & \dot\gam/2 \\
      -\dot\gam/2 & 0
    \end{array}\right),
\end{equation*}
there is a nonvanishing current manifested through the ever tumbling dumbbell
superficially indicating nonequilibrium. On the other hand, the length $|\x|$
undergoes only equilibrium fluctuations. For all practical purposes, this is
an equilibrium system. In fact, the applied power~\eqref{eq:work}
vanishes. Combining this result with the moving trap case discussed at the
beginning, we conclude that in the presence of flow the proper frame-invariant
criterion for distinguishing equilibrium from genuine nonequilibrium is
whether or not the applied power obeys $\dot w=0$.

Summarizing further, we have shown that considering external flow in the
expression for the work~\eqref{eq:work} [and therefore in the derived
quantities like dissipated heat~\eqref{eq:heat} and entropy
production~\eqref{eq:stot}] guarantees frame invariance of stochastic
thermodynamics. While formally two expressions for the asymmetric part of the
path probability under time reversal are possible which both fulfill the
fluctuation theorem, only one leads to a consistent identification of
thermodynamic quantities for the dynamics of soft matter systems in flow.

We thank R.~Finken and R.~K.~P.~Zia for inspiring discussions.


\end{document}